\documentclass{aa}
\usepackage{psfig,natbib} 
\usepackage{txfonts}
\def\feka{Fe K$\alpha$}
 
\def\xmm{{\it XMM-Newton}}

\def\rxte{{\it RXTE}}

\def\rosat{{\it ROSAT}}

\def\3c{3C 390.3}
\def\ltsima{$\; \buildrel < \over \sim \;$}
\def\simlt{\lower.5ex\hbox{\ltsima}} 
\def\gtsima{$\; \buildrel > \over \sim \;$}
\def\simgt{\lower.5ex\hbox{\gtsima}} 

\begin{document}
\title{Correlated spectral and temporal changes in 3C~390.3: \\
a new link between AGN and Galactic Black Hole Binaries?}
\author{M. Gliozzi\inst{1}
\and I.E. Papadakis\inst{2}
\and  C. R\"ath\inst{3}
}
\offprints{mario@physics.gmu.edu}
\institute{George Mason University, Department of Physics \&
Astronomy \& School of Computational Sciences, 4400 University Drive, 
MS 3F3, Fairfax, VA 22030
\and Physics Department, University of Crete, 710 03 Heraklion,
Crete, Greece
 \and   Max-Planck-Institut f\"ur 
extraterrestrische Physik, Postfach 1312, D-85741 Garching, Germany             }

\date{Received ...; accepted ...}

\abstract{This work presents the results from a systematic search for evidence
of temporal changes (i.e., non-stationarity)  associated with spectral
variations in \object{3C~390.3}, using data from a two-year intensive RXTE
monitoring campaign of this broad-line radio galaxy. In order to exploit the
potential information contained in a time series more efficiently, we adopt a
multi-technique approach, making use of linear and non-linear techniques. All
the methods show suggestive evidences for non-stationarity in the temporal
properties of 3C~390.3 between 1999 and 2000, in the sense that the
characteristic time-scale of variability decreases as the energy spectrum of
the source softens. However, only the non-linear, ``scaling index method" is
able to show conclusively that the temporal characteristics of the source do
vary, although the physical interpretation of this result is not clear at the
moment. Our results indicate that the variability
properties of 3C~390.3 may vary with time, in the same way as
they do in Galactic black holes in the hard state, strengthening the
analogy between the X-ray variability properties of the two types of object.
 This is the first time that such a behavior is detected in an AGN X-ray light
curve. Further work is needed in order to investigate whether this is a common
behavior in AGN, just like in the Galactic binaries, or not.
\keywords{Galaxies: active -- Galaxies:
  nuclei -- X-rays: galaxies } }
\titlerunning{Correlated spectral and temporal changes in 3C~390.3}
\authorrunning{M.~Gliozzi et al.}
\maketitle
%

\section{Introduction}

Among the basic properties characterizing an Active Galactic Nucleus
(AGN) the flux variability, especially at X-rays, is one of the most
common.  Previous multi-wavelength variability studies (e.g., Edelson et
al. 1996; Nandra et al. 1998) have shown that AGN are variable in every
observable wave band, but the X-ray flux exhibits variability of larger
amplitude and on time-scales shorter than any other energy band,
indicating that the X-ray emission originates in the innermost regions of
the central engine.

However, even though the discovery of X-ray variability dates back more
than two decades, its origin is still poorly understood. Nevertheless, it
is widely acknowledged that the variability properties of AGN are an
important means of probing the physical conditions in their emission
regions.  The reason is that using energy spectra alone it is often
impossible to discriminate between competing physical models and thus the
complementary information obtained from the temporal analysis is crucial
to break the spectral degeneracy.

Over the years, several variability models have been proposed, involving
one or a combination of the fundamental components of an AGN: accretion
disk, corona, and relativistic jet. These models can be divided into two
main categories: 1) intrinsically linear models, as the shot noise model
(e.g., Terrel 1972), where the light curve is the result of the
superposition of similar shots or flares produced by many independent
active regions, and 2) non-linear models, which require some coupling
between emitting regions triggering avalanche effects, as the
self-organized criticality model (e.g., Mineshige et al.1994), or models
assuming that the variability is caused by variations in the accretion
rate propagating inwards (e.g. Lyubarskii 1997).  The ``rms-flux
relation'' recently discovered by Uttley \& McHardy (2001), along with
detection of non-linearity in different AGN light curves (i.e., in 3C~345
by Vio et al. 1992; in 3C~390.3 by Leighly \& O'Brien 1997; in NGC~4051
by Green et al. 1999; in Ark~564 by Gliozzi et al. 2002), favors
non-linear variability models, but there is no general consensus on the
nature of the X-ray variability yet.

Because of their brightness, the temporal and spectral properties of
Galactic black holes (GBHs) are much better known and can be used to
infer information on their more powerful, extragalactic analogs, the AGN.
It is well established that GBHs undergo state transitions (see
McClintock \& Remillard 2003 for a recent review), switching between
``low'', ``intermediate", ``high'', and ``very high" states, which are all
unambiguously characterized
by a specific combination of energy spectrum and power spectral density
(PSD). Importantly, even when in the same state, the timing properties of
GBHs may vary with time, often accompanying changes of the spectral
properties. For example, Pottschmidt et al. (2003) showed that changes in
the PSD of Cyg X-1 are associated with changes in the slope of the energy
spectrum. Specifically, during the ``low/hard state'', the PSD
characteristic frequencies increase, as the spectral slope becomes
steeper.

\begin{figure*}[th]
\psfig{figure=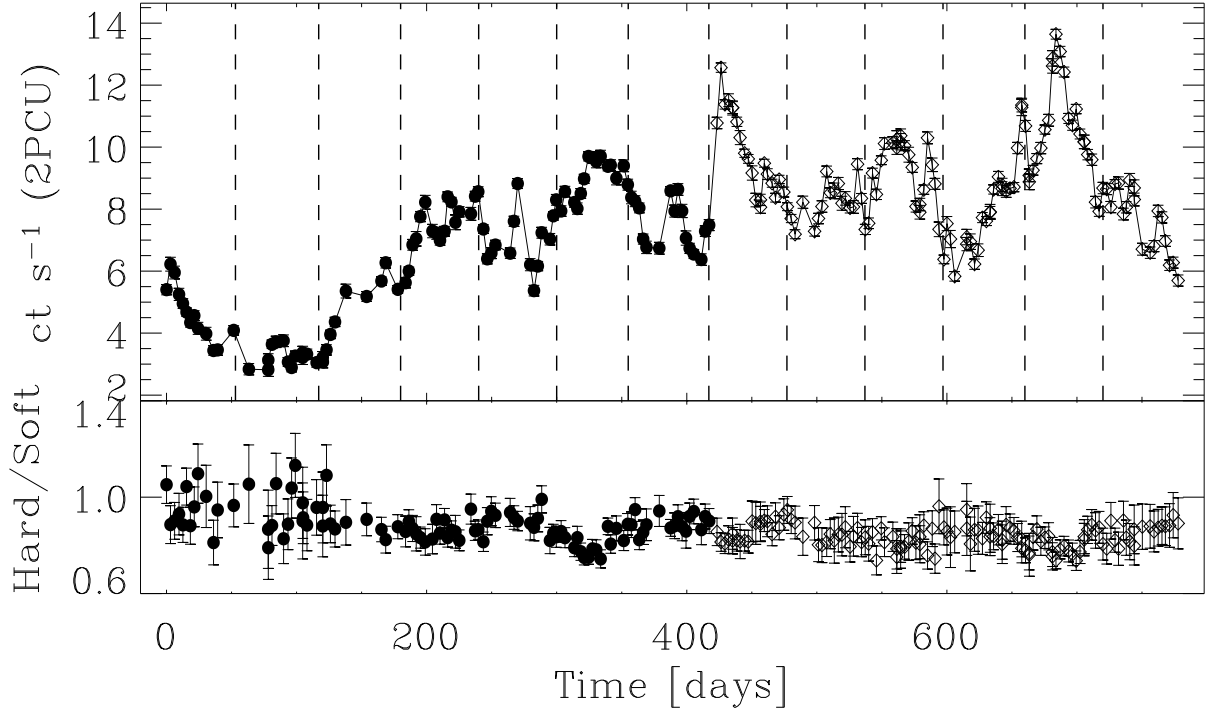,height=10cm,width=16.5cm,%
bbllx=32pt,bblly=6pt,bburx=530pt,bbury=290pt,angle=0,clip=}
\caption{X-ray light curves of the background-subtracted count rate in
the 2--20 keV band (top panel) and of the X-ray color 7--20 keV/2--5
keV (bottom panel) from RXTE PCA observations of 3C 390.3.  Time bins
are 5760s ($\sim$ 1 RXTE orbit). Filled circles represent data points
from the first monitoring campaign from 1999 January 8 to 2000
February 29, whereas open diamonds are data points from the second
monitoring campaign from 2000 March 3 to 2001 February 23. The
dashed lines in the top panel indicate the intervals used for the
time-resolved spectral analysis (see $\S$9).}
\label{figure:lc}
\end{figure*}

Transitions between ``different'' states have not been observed yet in
AGN. The reason may be associated with the fact that their flux
variability is characterized by much longer time-scales. Nevertheless, a
correspondence between GBH spectral states and some AGN classes has been
hypothesized. Specifically, Broad-Line Seyfert 1 galaxies have been
associated with the GBH low/hard state, and Narrow-Line Seyfert 1
galaxies with the GBH high/soft state, (for a recent discussion on this
topic see, for example, McHardy et al. 2004). 

Although genuine spectral transitions in AGN are unlikely to be detected
due to the longer time-scales involved, it is perhaps possible to detect
changes in the temporal properties (i.e., loss of stationarity)
associated with spectral changes, similarly to what observed in GBHs
within the same spectral state. To this end, of crucial importance are
the numerous monitoring campaigns carried out by \rxte\ in the last
decade, which have produced data sets suitable for searching for
non-stationarity in the AGN light curves, where long-term spectral
changes proved to be ubiquitous.

In this work, we present the results from a systematic search for evidence of
temporal changes associated with significant spectral variations in the case of
the broad-line radio galaxy (BLRG) \object{3C~390.3}.
We use data from a two-year intensive \rxte\
monitoring campaign.  Previous X-rays studies with \rosat, and \rxte\
(Leighly \& O'Brien 1997, Gliozzi et al. 2003) demonstrated that \object{3C~390.3}
is one of the most variable AGN on time-scales of days and months. Our motivation
stems from the results of a long monitoring campaign carried out with 
\rosat\ HRI, which showed evidence for possible non-stationarity
(and non-linearity) in the soft X-ray light curve of \object{3C~390.3} (Leighly \& O'Brien 1997).

\section{Observations and Data Reduction}
We use archival \rxte\ data (PI: Leighly) of 3C~390.3.  This source
was observed by \rxte\ for two consecutive monitoring campaigns
between 1999 and 2001.  The first set of observations was carried out
from 1999 January 8 to 2000 February 29, and the second one from 2000
March 3 to 2001 February 23.  Both campaigns were performed with
similar sampling: 3C~390.3 was regularly observed for $\sim$
1000--2000 s once every three days.  The observations were carried out
with the Proportional Counter Array (PCA; Jahoda et al. 1996), and the
High-Energy X-Ray Timing Experiment (HEXTE; Rotschild et al. 1998) on
\rxte. Here we will consider only PCA data, because the
signal-to-noise of the HEXTE data is too low for a meaningful timing
analysis.

The PCA data were screened according to the following acceptance
criteria: the satellite was out of the South Atlantic Anomaly (SAA) for
at least 30 minutes, the Earth elevation angle was $\geq 10^{\circ}$, the
offset from the nominal optical position was $\leq 0^{\circ}\!\!.02$, and
the parameter ELECTRON-2 was $\leq 0.1$. The last criterion removes data
with high particle background rates in the Proportional Counter Units
(PCUs). The PCA background spectra and light curves were determined using
the ${\rm L}7-240$ model developed at the \rxte\ Guest Observer Facility
(GOF) and implemented by the program {\tt pcabackest} v.2.1b. Since the
two monitoring campaigns span three different gain epochs, the
appropriate files ($pca\_bkgd\_cmfaintl7\_e3(4,5)bv20031123.mdl$,
depending on the observation
date), provided by the \rxte\ Guest Observer Facility (GOF), were used to
calculate the background light curves. This model is appropriate for
``faint'' sources, i.e., those producing count rates less than 40 ${\rm
s^{-1}~PCU^{-1}}$. All the above tasks were carried out using the {\tt
FTOOLS} v.5.1 software package and with the help of the {\tt rex} script
provided by the \rxte\ GOF, which also produces response matrices and
effective area curves appropriate for the time of the observation. Data
were initially extracted with 16 s time resolution and subsequently
re-binned at different bin widths depending on the application.  The
current temporal analysis is restricted to PCA, STANDARD-2 mode, 2--20
keV, Layer 1 data, because that is where the PCA is best calibrated and
most sensitive. Since PCUs 1, 3, and 4 were frequently turned off (the
average number of PCUs is 2.25 and 1.95 in the 1999 and 2000
observations, respectively), only data from the other two PCUs (0 and 2)
were used. All quoted count rates are for two PCUs.

The spectral analysis of PCA data was performed using the {\tt XSPEC
v.11} software package (Arnaud 1996). We used PCA response matrices and
effective area curves created specifically for the individual
observations by the program {\tt pcarsp}, taking into account the
evolution of the detector properties.  All the spectra were rebinned so
that each bin contained enough counts for the $\chi^2$ statistic to be
valid. Fits were performed in the energy range 4--20 keV, where the
signal-to-noise ratio is the highest.

\section{The X-ray Light Curve}

Figure~\ref{figure:lc} shows the background-subtracted 2--20 keV count
rate and hardness ratio (HR=7--20 keV/2--5 keV) light curves of 3C~390.3.
Filled circles represent data points from the first monitoring campaign,
whereas open diamonds are used for the second campaign.  A visual
inspection of the HR and count rate light curves indicates the presence
of a typical Seyfert-like spectral evolution (the energy spectrum softens
when the source brightens), and suggests the existence of different
variability patterns during the two monitoring campaigns. In the first
one, 3C~390.3 exhibits an initial smooth decrease in the count rate (a
factor $\sim$2 in 100 days) followed by an almost steady increase (a
factor $\sim$3.5 in 100 days). After day $\sim$ 200 (from the beginning
of the observation), the source shows significant, low-amplitude, fast
variations. During the second monitoring campaign, the source does not
show any long-term, smooth change in count rate, but only a flickering
behavior with two prominent flares. This qualitative difference is
associated with marked differences in the two mean count rates,
$6.4\pm0.2{~\rm s^{-1}}$ and $9.0\pm0.1{~\rm s^{-1}}$, and in the
fractional variability amplitudes, $F_{\rm var,99}=(33\pm3)$\% and
$F_{\rm var,00}=(19\pm2)$\%, calculated over each campaign
(see, e.g., Vaughan et al. 2003 for a detailed description of $F_{\rm var}$
and the estimate of its uncertainty). The
uncertainties have been computed assuming that the data distribution is
normal and the data are a collection of independent measurements.
However, the data points of AGN light curves are not independent but
correlated and thus the apparent difference between the mean count rates
and $F_{\rm var}$ could simply result from the red-noise nature of the
variability process, combined with its randomness (see, e.g., Vaughan et
al. 2003).

\section{Stationarity Tests}
Almost all methods employed in timing analysis (e.g., power spectrum,
structure function, auto-correlation analysis) assume some kind of
stationarity. Generally speaking, a system is considered stationary, if
the average statistical properties of the time series are constant with
time. More specifically, the stationarity is defined as ``strong'' if all
the moments of the probability distribution function are constant, and
``weak'' if only the moments up to the second order (i.e., mean and
variance) are constant. Usually, non-stationarity is considered an
undesired complication. However, in some cases the non-stationarity
provides the most interesting information, as in the case of GBHs
switching from a spectral state to another.

In previous temporal studies of AGN, no strong evidence for
non-stationarity has been reported. Indeed, AGN light curves are
considered to be ``second-order'' stationary, due to their red-noise
variability. Here, we adopt several complementary methods to investigate
the issue of the stationarity in the \object{3C~390.3} light curve.  
First, the temporal properties are studied with traditional linear
techniques in the Fourier domain, by estimating the PSD, and in the time
domain with the structure function analysis. Secondly, we probe directly
the probability density function associated with the time series.
Finally, we utilize the scaling index method, a technique borrowed from
non-linear dynamics, useful for detecting weak signals in noisy data.

All the above methods are separately applied to the 1999 and 2000
monitoring campaigns. We quantify the differences between the estimated
functions (like the power spectrum, the structure function, etc.) using
appropriate ``statistics". The probability that these statistics will
have the observed values are rigorously estimated with extensive Monte
Carlo simulations. These simulations allow us to assess whether the
observed values are due to measurement uncertainties (i.e., Poisson
noise), to the light curve sampling, to the red-noise nature of the
variability process, or to intrinsic variations of the statistical
properties of the source.

\subsection{Analysis in the Fourier Domain: Power Spectral Density}

We followed the method of Papadakis \& Lawrence (1993) to compute the
power spectrum of the 1999 and 2000 light curves, after normalizing them
to their mean. We used a 3-day binning (i.e., equal to the typical
sampling rate of the source) for both light curves. As a result, the
light curves that we used are evenly sampled, but there are a few missing
points in them; 30 out of 140 and 15 out of 120 for the 1999 and 2000
light curves, respectively. These points are randomly distributed over
the whole light curve, and we accounted for them using linear
interpolation between the two bins adjacent to the gaps, adding
appropriate Poisson noise.

\begin{figure}
\psfig{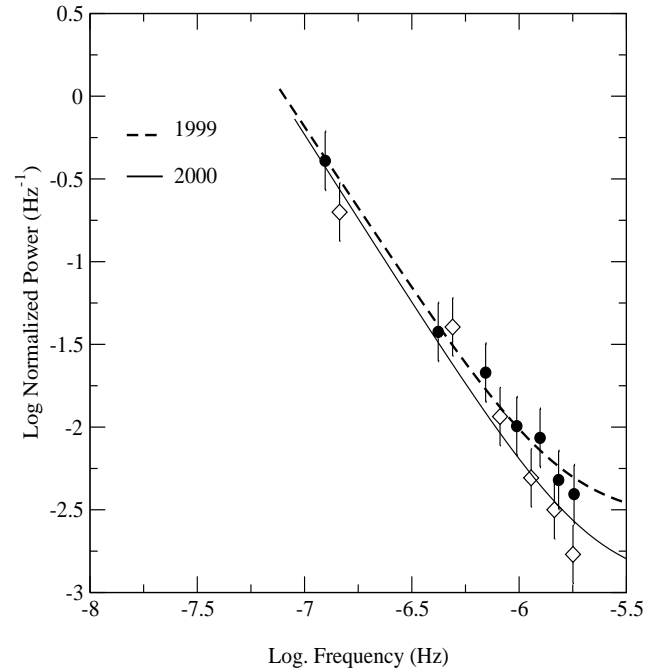}
\caption{RXTE PCA power density spectra using based on the 1999 and
2000 light curves with 3-day bins, normalized to their mean (filled
circles and open diamonds, respectively). The solid and dashed lines
show the best fitting power law models, including the effects of the
experimental Poisson noise level.}
\label{figure:psd}
\end{figure}

Figure~\ref{figure:psd} shows the PSD of the 1999 and 2000 light curves
(filled circles, and open diamonds, respectively). Each point in the
estimated PSD represents the average of 10 logarithmic periodogram
estimates.  We fitted both spectra with a simple power law ($P(f)\propto
f^{-a}$).  The model provides a good description of both PSDs; the best
fitting slope values (shown in Fig~\ref{figure:psd}) are consistent
within the errors with the value of $2\pm 0.3$ in both cases.

A closer look at the PSD plot indicates a possible low frequency
flattening of the 2000 PSD between $\log(\nu)=-6.5$ and $-7$,
corresponding to a time interval between $\sim$35 and 115 days.  In order
to investigate this issue further we split the two lowest frequency
points into two (to increase the frequency resolution) and refitted the
2000 PSD with a broken power-law model. We kept the slope below and above
the frequency break fixed to the values $1$ and $2$, respectively, and
kept as free parameters the break frequency and normalization of the PSD.
The model provides a good fit to the 2000 PSD, but statistically not
better than a simple power law according to an F-test. The best fitting
break frequency is $2.5\times10^{-7}$ Hz, which corresponds to a break
time-scale of $\sim 46$ days. The $90$\%-confidence lower limit to this
time-scale is 33 days, while, due to the limited coverage of the observed
PSD at lower frequencies, the upper limit is undefined. We also tried to
fit the 1999 PSD with this model, but without success: the best fitting
break frequency turned out to be lower than the lowest frequency sampled
by the 1999 light curve, indicating that there is no indication of any
breaks in the 1999 PSD. This power spectrum follows a power-law shape
with slope $a=2$ down to the lowest sampled frequencies.

In order to quantify the comparison between the 1999 and 2000 PSDs in a
statistical way, we used the ``{\it S}'' statistic as defined by
Papadakis \& Lawrence (1995). To avoid the constant Poisson noise level,
we considered the periodogram estimates up to frequencies $\sim 7\times
10^{-7}$ Hz. We find that $S=-1.3$, which implies that the two light
curves are not far away from the hypothesis of stationarity. Therefore,
the differences suggested in the previous paragraph by the broken
power-law model fitting results are not supported by this more rigorous
statistical test.  Light curves with denser sampling, which would result
in a much higher frequency resolution at lower frequencies (where the
potential differences between the two PSDs are observed), would be
necessary for a more sensitive comparison between the 1999 and 2000 power
spectra.

Recent results from the power spectrum analysis of combined \xmm\ and
\rxte\ light curves of radio-quiet AGN (Markowitz et al., 2003; McHardy
et al. 2004) show that the $2-10$ keV power spectrum of AGN follows a
power-law form with a slope of $2$ down to a ``break'' frequency, below
which it flattens to a slope of $1$. This break time-scale appears to
scale with the black hole mass according to the relation: $t_{br}(\rm
days)=M_{\rm BH}/10^{6.5}~{\rm M}_{\odot}$ (Markowitz et al. 2003). Using
the reverberation mapping technique, Peterson et al. (2004)  estimate the
black hole mass of 3C~390.3 to be $M_{\rm BH}\sim 2.9\times 10^{8}$
M$_{\odot}$.  In this case, the characteristic break time-scale of
\object{3C~390.3} should be $\sim 90$ days, which corresponds to a break
frequency of $\sim 10^{-7}$ Hz.  This break frequency is entirely
consistent with the broken power-law model fitting results in the case of
the 2000 light curve. Unfortunately, though, even the combined 1999 and
2000 PSD does not extend to frequencies low enough for this break
frequency to be firmly detected.
 
It is worth noticing that the value assumed for the mass of the black
hole harbored by \object{3C~390.3} is still a matter of debate. The
value used in this work is the most recent and it is consistent within
the errors with the mass estimate given by Kaspi et al. (2000).
However, based on another reverberation mapping experiment on this
source, Sergeev et al. (2002) find a longer reverberation lag and
determine a higher mass of $2\times10^9$ M$_{\odot}$ using a different
method. In that case, the putative break frequency in the PSD would be
located at a frequency of $\sim$ 1/yr, which could not be probed with
the current data set.

We conclude that the combined 1999+2000, $2-20$ keV PSD of 3C~390.3 is
well fitted by a simple power-law model at all sampled time-scales from
$3^{-1} - 100^{-1}$ days$^{-1}$. The same model fits well the individual
1999 and 2000 PSDs as well. There is an indication of a low-frequency PSD
break (at a frequency which corresponds to a time-scale of $\sim$50 days)
in the 2000 PSD, which is absent in the 1999 PSD. However, due to the low
resolution of the power spectra at hand, we cannot be certain of the
significance of the difference between the two PSDs.

\subsection{Analysis in the Time Domain: Structure Function}

Next, we performed an analysis based on the structure function (e.g.,
Simonetti 1985), a linear method which works in the time domain and has
the ability to discern the range of time-scales that contribute to the
variations in the data set.  In principle, the SF should provide the same
information as the PSD, as both functions are related to the
auto-covariance function of the process $R(\tau)$. The power spectrum is
defined as the Fourier transform of $R(\tau)$, while the intrinsic SF at
lag $\tau$ is equal to $2\sigma^{2}-R(\tau)$, where $\sigma^{2}$ is the
variance of the process (Simonetti et al., 1985). In fact, if the PSD
follows a power law of the form $P(f)\propto f^{-a}$ (where $f$ is the
frequency), then SF $\propto \tau^{a-1}$ (Bregman et al., 1990).

\begin{figure}
\psfig{figure=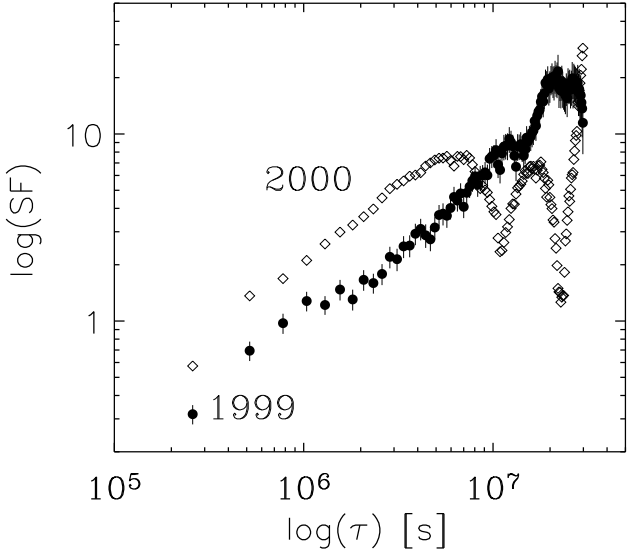,width=8.7cm,%
bbllx=60pt,bblly=60pt,bburx=350pt,bbury=320pt,angle=0,clip=}
\caption{Structure function of \object{3C~390.3} during
the 1999 (filled circles) and 2000 (open diamonds) monitoring
campaign. Time bins are 3 days. For sake of clarity, error-bars were plotted only
for the 1999 SF.}
\label{figure:SF}
\end{figure}

We computed the structure functions for the first and second monitoring
campaigns. The results are plotted in Fig~\ref{figure:SF}: the 
filled circles refer to the 1999 light curve, whereas the 
open diamonds describe the 2000 SF. Both SFs have a power-law shape of
the form SF $\propto \tau^1$, which is consistent with the results of the
PSD analysis, since $P(f)\propto f^{-2}$ (see \S{4.1}). Furthermore, the
2000 SF shows a flattening to SF $\propto \tau^0$, at a characteristic
time lag of $\sim 4-6\times 10^{6}$ s, followed by an oscillating
behavior. This value corresponds to a characteristic time-scale of $\sim
50-70$ days, which is fully consistent with the putative break suggested
by the 2000 PSD and corresponds approximately to the value expected on
the basis of the black hole mass estimate (see discussion on \S{3.1}).

The 1999 and 2000 SFs look qualitatively different. However, the
statistical significance of this difference cannot be assessed directly,
since the the SF points are correlated and their true uncertainty is
unknown. The errors shown in Fig.~\ref{figure:SF} are representative of
the typical spread of the points around their mean in each time-lag bin,
hence they depend mainly on the the number of points that contribute to
the SFs estimation at each time lag. To investigate in a quantitative way
whether the difference between the two SFs can be simply ascribed to the
intrinsic, red-noise randomness of the light curves or it actually
reveals a non-stationary behavior, the use of Monte Carlo simulations is
necessary.

\begin{figure}
\psfig{figure=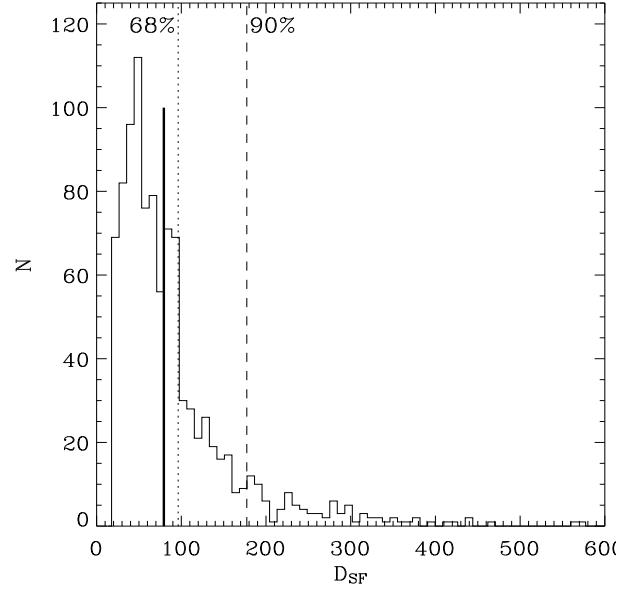,height=8cm,width=8.cm,%
bbllx=55pt,bblly=35pt,bburx=410pt,bbury=410pt,angle=0,clip=}
\caption{Comparison between $D_{\rm SF,obs}$ (thick continuous
line) and the distribution of $D_{\rm SF,synth}$ obtained with
red-noise simulations.}
\label{figure:DSF}
\end{figure}

For this reason, we performed a model-dependent numerical experiment assuming
that both the 1999 and 2000 light curves are realizations of a process with
mean count rate equal to the average value of the 1999 and 2000 light curves,
and with a power spectrum of power-law index $2$ down to a break  frequency 
below which the power-law index changes to $1$. We performed the experiment
considering various PSD break time scales:  $t_{\rm br}= 50, 100, 200,$ and 300
d. Here we present the  only the results from $t_{\rm br}=$300 d because they
provide the most conservative limits on the significance of the observed
differences. In all cases, the PSD normalization was fixed in such a way that
the model PSD will have the same value as the observed 1999+2000 best fitting
power law at 10$^{-6}$ Hz.  Assuming this power spectrum model, we created two
sets of $10^3$ synthetic light curves  (using the method of Timmer \& Koenig
1995). Importantly, for each synthetic light curve, the appropriate \rxte\
background light curve (as provided by the {\small FTOOL PCABACKEST}) was added
to each light curve, and then the appropriate Poisson noise was added to each
point in them.  The length of the synthetic light curves was 20 times longer
than the length of the observed light curves, and their points  were separated
by intervals of 160 s. In this way, we could simulate the effects of the red
noise leakage in the estimation of the SF, although the aliasing effects were
not accounted for. This should not be a serious problem though, as the steep
power spectral slope (-2) implies that these effects should not be very
strong.  Each synthetic light curve   was then re-sampled in  such a way to
have the same length and sampling pattern as the observed light curves.

It is worth noticing that, after fixing the PSD normalization, the
synthetic light curves are not renormalized to the observed variance, but
their variance is free to vary, allowing a larger scatter in the
low-frequency power and hence maximizing the effects of long-term flux
variations. For completeness, we also performed a second set of
``standard" red-noise simulations with synthetic light curves
renormalized in such a way to have the same sampling pattern, mean, and
variance as the 1999 and 2000 light curve, respectively. No significant
differences in the red-noise tests of SF are found using the two
different numerical approaches.

For each pair of synthetic light curves, we computed their SFs, and used
the sum of the squared ratios between them at each lag, i.e.  
$D_{SF}=\sum[SF_2(\tau_i)/SF_1(\tau_i)]^2$, to quantify their difference.  
Based on the distribution of the $D_{\rm SF,synth}$ values (see
Fig.~\ref{figure:DSF}), we find that the probability that the $D_{\rm
SF,obs}$ value would be as large as observed, is of the order of 44\%,
Therefore, the apparent differences between the observed 1999 and 2000
SFs shown in Fig.~\ref{figure:SF} can be the result of the red-noise
character of the variability process.

However, we noticed that the synthetic SFs show the largest differences
at lags larger than $\sim 100$ days (i.e., at those lags where the
wiggles and bumps in the observed SF$_{00}$ appear). In fact, when we
restricted the computation of the $D_{SF}$ up to lag=90 days (thus
including the flattening of the observed $SF_{00}$, but excluding the part
which shows the highly erratic behavior at large lags) we found that
probability of the difference between the observed 1999 and 2000 SFs
being due to red-noise effect decreases to 7.5\%.
 
The SF analysis results are similar to those from the power spectrum
analysis. The $SF_{00}$ shows a flattening around lag $\sim 50-70$ days,
which is absent in the $SF_{99}$, suggesting that the power spectrum
break may have indeed moved at higher frequencies during the 2000
observations, in agreement with the 1999 and 2000 PSDs. However, just
like with the observed PSDs, we cannot be certain of the reality of this
effect.

\subsection{Time-independent Analysis: Probability Density Function}

\begin{figure}
\psfig{figure=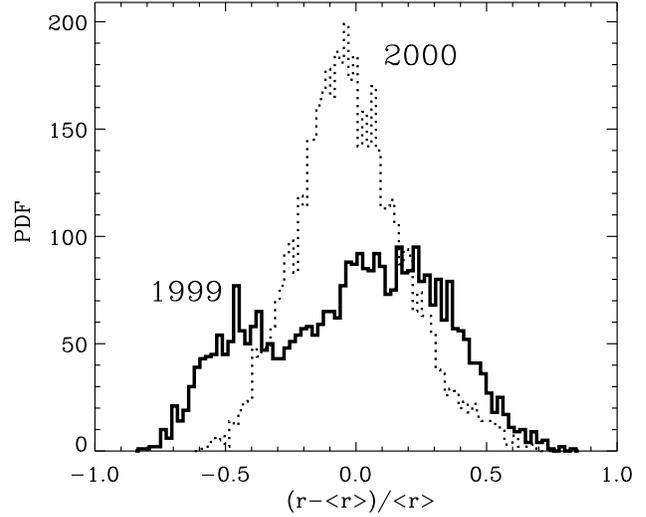,width=8.7cm,%
bbllx=40pt,bblly=60pt,bburx=365pt,bbury=325pt,angle=0,clip=}
\caption{Probability density function of count rates scaled and
normalized with respect to the average count rate $\langle r \rangle$,
during the 1999 (thick solid line) and 2000 (dotted line) monitoring campaign.  
The time bins are 32~s wide. }
\label{figure:PDF}
\end{figure}

After utilizing methods which are based on the temporal order of the data
points, here we investigate the stationarity of 3C~390.3 with a method
independent of the temporal order of the data points, the probability
density function (PDF), which is the histogram of the different count
rates recorded during the monitoring campaign.  Fig.~\ref{figure:PDF}
shows the normalized distributions $(r-\langle r \rangle)/\langle r
\rangle$ of the 1999 (thick solid line) and 2000 (dotted line) monitoring
campaigns.  The first transformation ($r-\langle r \rangle$) brings both
light curves to zero mean, the second, normalizes the amplitudes to unit
mean. In this way, the relative amplitudes of the light curves are
directly probed, and the PDF results can be directly compared to the
power spectral ones, which use exactly the same convention.

For a better visual comparison, we have used small time bins of 32~s in
Fig.~\ref{figure:PDF}. However, for a quantitative
comparison between PDFs (see below), we have used larger time bins to
prevent any contamination from the Poisson noise.  There is a clear
qualitative difference between the two PDFs, with the 1999 distribution
characterized by two broad peaks and the 2000 PDF having a bell-like
shape. This qualitative difference is confirmed by a Kolmogorov-Smirnov
(K-S) test, which gives a probability of 99.99\% that the two
distributions are different.

One may ask what are the effects on the PDF properties of the light
curve bin size, whether the PDF is really time independent (a question
that is directly related to the length of the light curve), and
whether the K-S test is appropriate to quantify the statistical
difference between two PDFs or not. The answers to these questions are
not clear. In all cases, the most important issue is the red-noise
character of the AGN light curves (in all wavebands). As the size of
the light curve bins decreases, the number of the points in the PDF
increases, and the sampled PDF is better defined. However, as the
points added are heavily correlated due to the red-noise character of
the variations, no ``extra'' information is added to the PDF. On the
other hand, since the ``sensitivity'' of the K-S test increases with
the number of points in the sampled PDF (which are assumed to be
independent), there is a possibility that differences which are purely
due to the stochastic nature of a possible stationary process will be
magnified and the K-S test will erroneously indicate ``significant''
differences between two PDFs. Finally, one should also consider the
effects of the Poisson noise to the sampled PDFs. In the cases where
the source signal has a much larger amplitude than the variations
caused by the experimental noise (as is the case here), we should not
expect this effect to be of any significance. In any case, the Poisson
noise should produce similar PDFs (once they are normalized to their
mean) so any significant differences cannot be due to this effect.

\begin{figure}
\psfig{figure=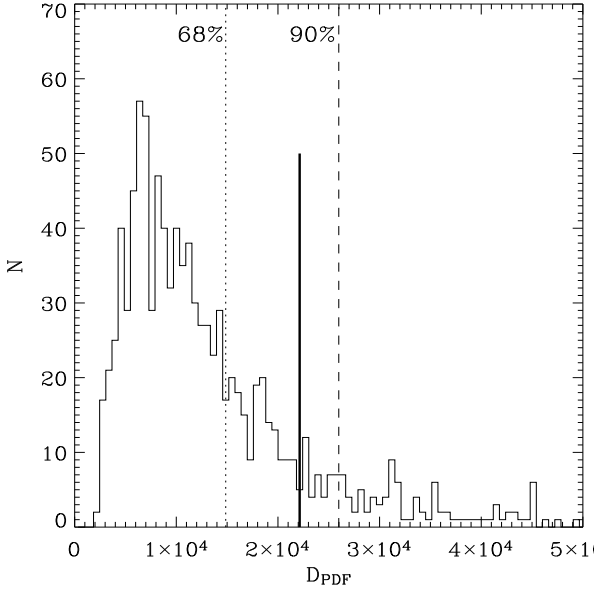,height=8cm,width=8.cm,%
bbllx=55pt,bblly=35pt,bburx=410pt,bbury=410pt,angle=0,clip=}
\caption{Comparison between $D_{\rm PDF,obs}$ (thick continuous
line) and the distribution of $D_{\rm PDF,synth}$ obtained with
red-noise simulations.}
\label{figure:DPDF}
\end{figure}

To assess the apparent PDFs differences shown in Fig.~\ref{figure:PDF}, in a
more rigorous way, we have used the $10^3$ synthetic light curves with
tbin=160 s, obtained from the Monte Carlo simulations described in $\S$4.2.
For each pair of synthetic PDFs, we computed the sum
$D_{PDF}=\sum[PDF_1(i)-PDF_2(i)]^2$, in order to quantify their differences. A
comparison of the distribution of the 10$^3$ synthetic $D_{\rm PDF,synth}$ with
the observed value $D_{\rm PDF,obs}$ (represented by a thick solid line in
Fig.~\ref{figure:DPDF}) reveals that a value as large as $D_{\rm PDF,obs}$
appears in 14.3\% of all the synthetic light curves. Consequently, the apparent
difference between PDF$_{99}$ and PDF$_{00}$ could be caused by red-noise
effects.

Looking closer at the synthetic PDFs that show the largest D$_{PDF}$
values, we noticed that they are caused mainly by differences at
$|(r-\langle r \rangle)/\langle r\rangle|>0/6$, where there is only a
small fraction of the total number of points in the histograms. For that
reason, we divided the x-axis into three sub-intervals (-1/-0.6;
-0.6/0.6; and 0.6/1) and calculated D$_{PDF}$ in each one of them. When
we consider the middle interval, the probability that the difference
between the observed PDFs can be explained to red-noise effects is
reduced to 12.6\%. 

In summary, the 1999 PDF appears to be broader than the 2000 PDF, which shows a
well defined, uniform distribution around its mean. It shows a bimodal shape,
with a large fraction of points having a count rate smaller than average. This
is qualitatively consistent with the results from the previous sections. For
example, in the case of a PSD break at low frequencies, the observed variations
are dominated by the components with long periods. If the PDF is estimated with
data sets shorter than the PSD break time-scale, we would expect the points to
show a broad distribution around the light curve mean, as in the case of the
1999 PDF.  Indeed, our results in the case of a PSD break at 1/50 days$^{-1}$
show that the observed 1999 PDF  is entirely inconsistent with this assumption,
contrary to the observed 2000 PDF which is consistent.  However, as before with
the results from the PSD and SF analysis, a red-noise process with a PSD break
at 1/300 days$^{-1}$ is in rough agreement with both the 1999 and 2000 PDFs.
So, we conclude that observed 1999 and 2000 PDF differences possibly support
the hypothesis of non-stationarity, but not at a high significance level.

\subsection{Non-linear Analysis: Scaling Index Method}

Non-linear methods are rarely employed in the analysis of AGN light
curves, partly because these methods are less developed than the
linear ones, partly because it is implicitly assumed that AGN light
curves are linear and stochastic. However, non-linear methods can be
useful not only to characterize the nature of chaotic deterministic
systems, but also to discriminate between two time series, regardless
of the fact that they are linear or non-linear. 

When methods of non-linear dynamics are applied to time series analysis,
the concept of phase space reconstruction represents the basis for most
of the analysis tools. In fact, for studying chaotic deterministic
systems it is important to establish a vector space (the phase space)
such that specifying a point in this space specifies a state of the
system, and vice versa.  However, in order to apply the concept of phase
space to a time series, which is a scalar sequence of measurements, one
has to convert it into a set of state vectors.

For example, let us suppose we have an observed time series, $x(t_{i})$,
with $i=\Delta t, 2\Delta t, ..., N\Delta t$. One can construct a set of
2(3, 4, ...)-dimensional vectors by selecting pairs (triplets,
quadruplets, ...) of data points, whose count rate values represent the
values of the different components of each vector. The data points
defining the vector components are chosen such that the second point is
separated from the first by a time delay $\Delta t'$, while the third is
separated from the first by $2 \Delta t'$, and so on. This process,
called time-delay reconstruction, can be easily generalized to any
n-dimensional vector. The time delay $\Delta t'$ is usually much larger
than the sampling interval $\Delta t$ of the observed light curve. In
fact, in non-linear dynamics, where this technique is used to determine
fractal dimensions and discriminate chaos from stochasticity, the choice
of the time delay is not arbitrary, but should satisfy specific rules
(see Kantz \& Schreiber 1997 for a review).

The set of n-dimensional vectors, derived from the original light curve,
is then mapped (or embedded) into an n-dimensional vector space,
producing a phase space portrait, whose topological properties represent
the temporal properties of the corresponding time series.

In order to quantify the characteristics of a phase space portrait, a
widely used method is the correlation integral $C(R)$ (Grassberger \&
Procaccia 1983).  
An exhaustive description of the correlation integral method and
its application to X--ray light curves of AGN is given by Lehto et al.
(1993).
\begin{figure}
\psfig{figure=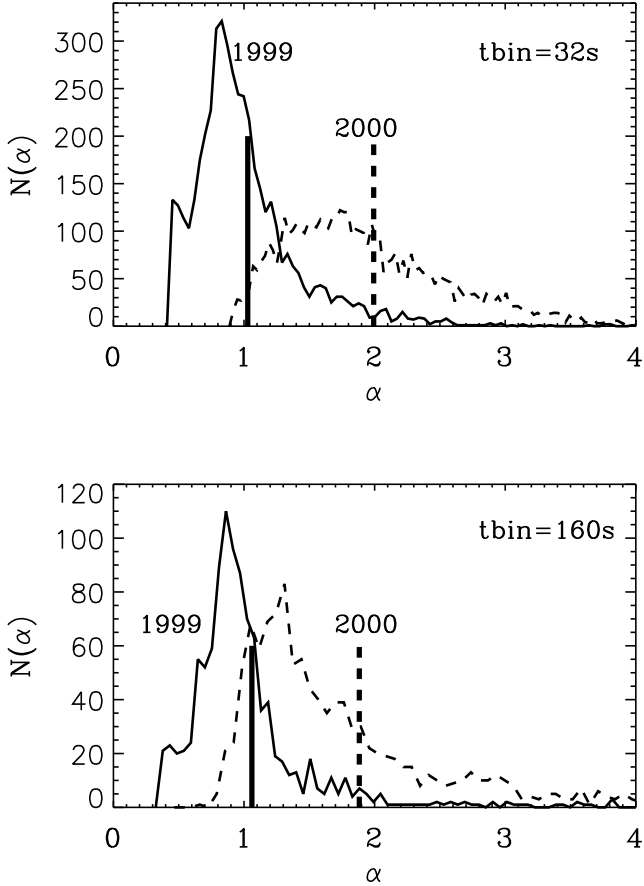,height=12cm,width=8.7cm,%
bbllx=50pt,bblly=35pt,bburx=405pt,bbury=515pt,angle=0,clip=}
\caption{Scaling index distributions of \object{3C~390.3} during the 1999
(solid line) and 2000 (dashed line) monitoring campaign, using light
curves binned at 32 s (top panel), and 160 s (bottom panel). The thick
continuous lines represent the scaling index mean value during 1999,
$\langle\alpha_{99}\rangle$, while the thick dashed lines represent
$\langle\alpha_{00}\rangle$.}
\label{figure:alpha}
\end{figure}

Another method that can quantify the topological properties of a phase
space portrait is the scaling index method (hereafter SIM; e.g.,
Atmanspacher et al. 1989)  which has been employed in a number of
different fields (e.g., R\"ath \& Morfill 1997; R\"ath et al. 2002),
including the study of AGN time series in the case of \object{Ark~564}
(Gliozzi et al. 2002).  Differently from the correlation integral, which
is a global indicator, the scaling index method measures both the global
and {\it local} properties of a phase space portrait.  Indeed SIM is
based on the local estimate of the correlation integral for each vector
in the phase space, by measuring the ``crowding'' of the vectors around
each individual point in the phase space.

For example, suppose we choose to define vectors of dimension $D$ from a
time series $x(t_i)$, with time delay $\Delta t'=\Delta t$. If the time
series has $N$ observed points, the number of vectors is $N'=N-D+1$.
Then, for each of the $N'$ vectors, say the $i-$th, we calculate the cumulative number
function,

\begin{equation} 
C_i(R)=n\{j\vert d_{ij}\leq R\}.
\end{equation}

In other words, $C_i(R)$ measures the number of vectors $j$ whose
distance $d_{ij}$ from the vector $i$ is smaller than $R$. 
For estimating the local scaling properties of the distribution of
vectors, one approximates the function $C_i(R)$ with a power law 
of the form $C_i(R)\sim R^{\alpha_i}$ within a given interval 
[$R_1,R_2$] (which is related
to the typical distances between the data points, that, in turn, depend
on the choice of the embedding space dimension). The exponent ${\alpha_i}$
is called {\it scaling index}.In practice, the scaling index for each
vector $i$ is computed by considering the logarithmic increase of $C_i(R)$
within a suitably chosen interval [$R_1,R_2$]. In all our calculations
${\alpha_i}$ has been determined as follows:

\begin{equation} 
{\alpha_i}=\frac{\log C_i(R_1,R_2)-\log C_i(R_1)}{(\log R_2-\log R_1)}.
\end{equation}
where $C_i(R_1,R_2)$ denotes the mean cumulative number within the
interval [$R_1,R_2$]. Note that, by definition, the scaling index cannot
have negative values.

The scaling index can be estimated for all the $N'$ vectors in the phase
space, producing a distribution of scaling indices, whose shape, width,
and centroid quantify the the topological properties of
the phase space portrait. In this way, the temporal properties of the
original light curve are translated into the properties of the scaling
index distribution. 

Note that for a purely random process 
(e.g., a process with an associated time series where the points vary
independently) the shape of the histogram is broad, dominated by the
right-ward tail, and the average scaling index $\langle\alpha\rangle$
tends to the value of the dimension of the embedding space. On the other
hand, for highly correlated processes and for deterministic (chaotic)
processes, the histogram is typically narrowly shaped, with the value of
$\langle\alpha\rangle$ always significantly smaller than the dimension of
the embedding space.

One of the most appealing characteristics of SIM is its ability to
discern underlying structures in noisy data. Therefore, it might be more
sensitive than linear methods like PSD, SF, and PDF, in identifying
intrinsic differences in the temporal properties of 3C~390.3 during 1999
and 2000.

To test the possible presence of two distinct temporal states of the system in
the 1999 and 2000 observations, we have estimated the distribution of scaling
indices of these two light curves. We stress that, at this point, we are not
interested in determining any kind of ``dimension"   of the system, since this
would be meaningless for non-deterministic systems. Our aim is to compare the
1999 and 2000 histograms of the scaling index values, and try to detect any
significant differences between them. To the extend that these histograms
correspond to the temporal properties of the system, any differences will imply
variations in the intrinsic temporal properties of the system.

We calculated the scaling index for all the vectors of the two phase
space portraits derived from the two monitoring campaigns, using
embedding spaces of dimensions ranging between $D=2$ and 10.  There are
no systematic prescriptions for the choice of the embedding dimension.
The discriminating power of the statistic based on the scaling index is
enhanced using high embedding dimensions. This can be understood in the
following way: if the data are embedded in a low-dimension phase-space
many of them will fall on the same position, losing part of
the information. On the other hand, the choice of too high embedding
dimensions would reduce the point density in the pseudo phase-space,
lowering the statistical significance of the test. As for the time delay
$\Delta t'$, we have used $\Delta t'=$32 s and 160 s, producing, with the
former choice, $\sim$4500--5000 vectors, and, with $\Delta t'=$160 s,
$\sim$1000--1200 vectors, depending on the dimension and on the campaign
(the 2000 campaign, albeit shorter, has longer individual observations, producing a
slightly higher number of data points and thus of vectors). It is worth
noticing that the use of small time bins cause the data points to be
correlated, but this is not necessarily a drawback, in particular in the
case where one wants to measure a difference in the degree of correlation
of two time-series.
\begin{figure*}[htb] \noindent
\psfig{figure=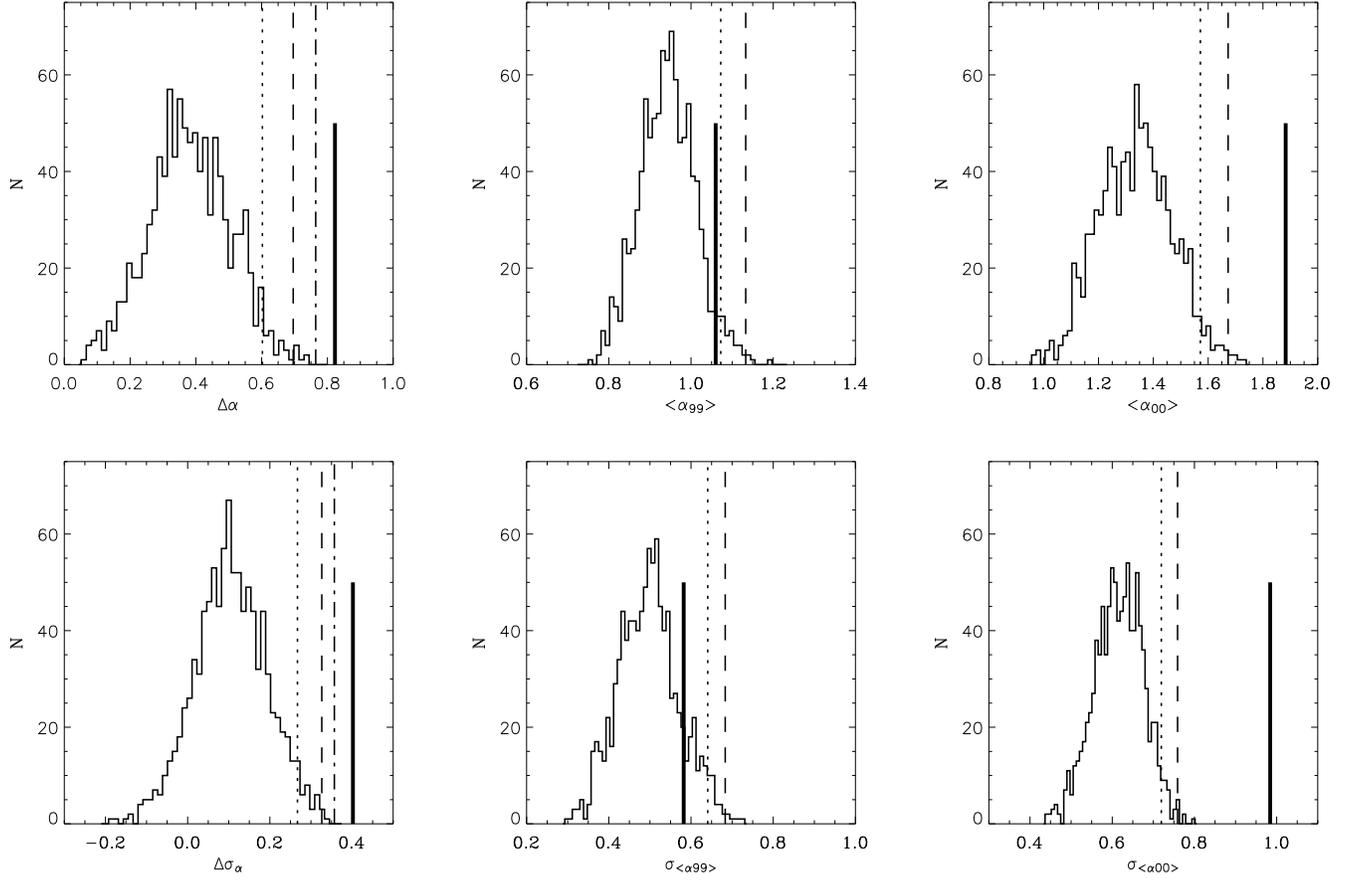,width=18cm,%
bbllx=37pt,bblly=200pt,bburx=537pt,bbury=537pt,angle=0,clip=} 
\caption{Top left panel: Distribution of $\Delta\alpha_{synth}$ obtained
from red-noise simulations (tbin= 160 s) compared with
$\Delta\alpha_{obs}$ (thick solid line). The dotted, dashed and
dot-dashed lines represent the 95\%, 99\%, and 99.9\% confidence levels,
respectively. Top middle and right panels: Distribution of mean synthetic
scaling indices obtained from red-noise simulations,
$\langle\alpha_{99(00),synth}\rangle$, compared with
$\langle\alpha_{99(00),obs}\rangle$ (thick solid line).  Bottom left
panel: Distribution of $\Delta\sigma_{synth}$, i.e., of the differences
between standard deviations around $\langle\alpha\rangle$ of the synthetic
distributions of scaling indices. Bottom middle and right panels:
Distribution of synthetic standard deviations around
$\langle\alpha\rangle$ in 1999 and 2000, respectively.  }
\label{figure:SIMrednoise2} 
\end{figure*} 
In all the cases (i.e. for all D values) we found that the 1999
scaling index histogram is systematically different from the 2000
histogram, with the former being significantly narrower, and its centroid
being smaller than the latter. This result suggests that the points in
the 1999 light curve are more ``strongly" correlated (e.g., over longer
time-scales) than those in the 2000 light curve.

Importantly, the results obtained using tbin=32 s are fully consistent
with those obtained with tbin=160 s, suggesting that the scaling index
method is actually able to detect the same underlying structure despite
the different levels of Poisson noise (by using tbin=32 s, the ratio
between the mean error and the mean count rate $\langle
err\rangle/\langle rate\rangle$ increases by a factor larger than 80\%
compared to tbin=160 s). The resulting histograms for a 5-dimensional
phase space for tbin=32 s (top panel) and tbin=160s (bottom panel) are
plotted in Fig.~\ref{figure:alpha}: the histograms with solid lines
represent data from the first monitoring campaign, whereas the histograms
with dashed lines refer to 2000. In both panels, the thick solid (dashed)
line represents the location of the centroid of $\langle\alpha\rangle$ in
1999 (2000), which remains roughly at the same location despite the large
change in the signal-to-noise level.  

\subsubsection{Impact of red and Poisson-noise effects on the SIM}

In the previous sections we have have seen how the red-noise effects
might affect the results from the SF and PDF analyses, casting doubts on
the statistical significance of apparently remarkable differences
observed in the temporal properties between 1999 and 2000. Therefore, it
is imperative to assess the impact of the red-noise effects on SIM.
\begin{figure*}[hbt]
\noindent
\psfig{figure=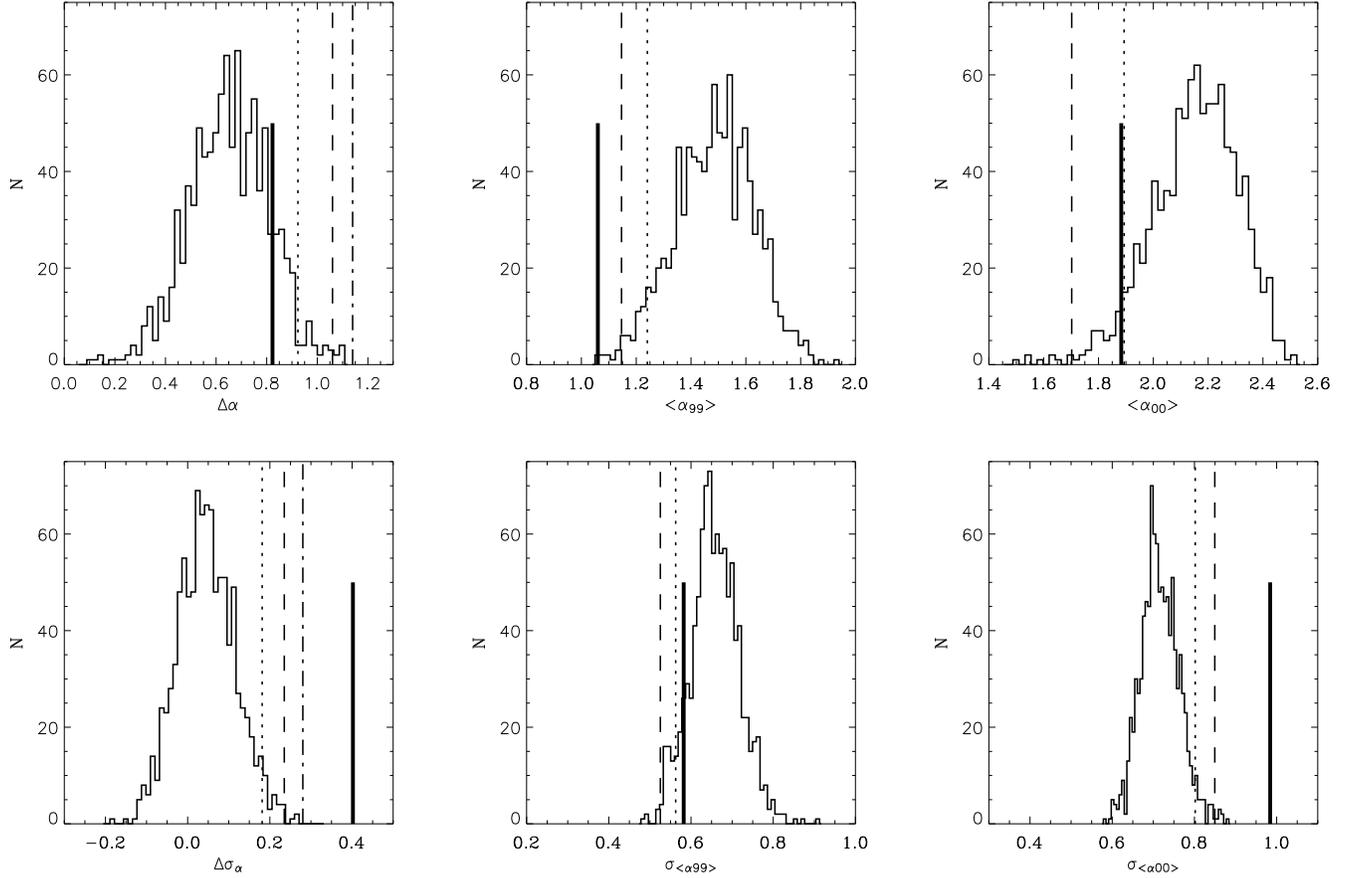,width=18cm,%
bbllx=37pt,bblly=200pt,bburx=537pt,bbury=537pt,angle=0,clip=}
\caption{Left panel: Distribution of $\Delta\alpha_{synth}$ obtained from
red-noise simulations with 50\% extra noise compared with
$\Delta\alpha_{obs}$ indicated by the thick continuous line (tbin= 160
s). Middle and right panels: Distribution of mean synthetic scaling
indices obtained from red-noise simulations with 50\% extra noise,
$\langle\alpha_{99(00),synth}$, compared with
$\langle\alpha_{99(00),obs}$ (thick continuous line).  The dotted and
dashed lines represent the 95\% and 99\% confidence levels, respectively.
}
\label{figure:SIMrednoise2b}
\end{figure*}

Before discussing the results of red-noise simulations, it is important to note
that in May 2000 (i.e., near the beginning of the second monitoring campaign of
3C~390.3) the propane layer on PCU0 was damaged, causing a systematic increase
of the background. This effect is accounted by the most recent PCA background
models and does not represent a problem when sufficiently large time bins are
used (http://lheawww.gsfc.nasa.gov/users/craigm/pca-bkg/bkg-users.html).
However, the SIM, being a statistical method, needs large amounts of points to
work properly and thus necessarily requires the use of small time bins. We
therefore paid particular attention at possible systematic effects caused by
the use of PCU0 data in 2000. Indeed, the SIM does notice a difference between
the PCU0 and PCU2 during 2000, with the former PCU being characterized by
larger values of $\langle\alpha\rangle$. To be on the safe side, we report the
SIM results obtained excluding the data from PCU0 in 2000. However, for the
sake of completeness, we also carried out all tests using both PCUs in 2000,
and reached the same conclusions.

To assess the impact of red-noise effects on SIM, we have used $10^3$
synthetic light curves obtained from Monte Carlo simulations very similar to
those described in \S4.2, the only difference being the use of one PCU (PCU2)
instead of two (PCU0+PCU2) for the simulated data of 2000. As before, we
report our results in the case when the PSD break at 1/300 day$^{-1}$.

For each pair of synthetic light curves the SIM is applied and the difference
$\Delta\alpha \equiv \langle\alpha_{00}\rangle - \langle\alpha_{99}\rangle$ is
computed. For simplicity and computational reasons (since we are dealing with
thousands of simulations of data-sets containing thousands of points), we limit
our discussion mainly to $\Delta\alpha$ and $\Delta\sigma_\alpha$ (i.e., the
difference between the dispersion of the $\alpha$ values around
$\langle\alpha\rangle$) as the discriminating parameters between the scaling
index histograms of the two light curves. However, it must be kept in mind that
SIM provides more information, and one can use the full distribution of the
scaling index values for the most accurate comparison between the two observed
histograms.

Fig.~\ref{figure:SIMrednoise2} (left panels) reveals that red-noise
effects are unable to explain $\Delta\alpha_{\rm obs}$ (top panel; thick
solid line) nor $\Delta\sigma_{\rm obs}$ (bottom panel) at very high
significance level (P$<0.1$\%).  It is worth noting that, if no
Poisson noise was added to the synthetic light curves, then the
distributions of $\Delta\alpha$ and $\Delta\sigma$ would be centered around
zero. The fact that the distributions are shifted towards positive values
is due to the use of two PCUs in 1999 and only one in 2000, and
hence to the different Poisson noise contributions to the 1999 and 2000
synthetic data.

To understand better these results, we have also plotted the
distributions of the scaling index centroids and standard deviations in
1999 and 2000 (Fig.~\ref{figure:SIMrednoise2} middle and right panels,
respectively). It is clearly shown that the distributions of
simulated data of 1999 (Fig.~\ref{figure:SIMrednoise2} middle  
panels) are marginally consistent with the corresponding 
observed values: the chance probabilities to obtain values consistent
with $\langle\alpha_{99}\rangle$ and $\sigma_{\alpha,99}$ are P$\sim 6$\% 
and P$\sim 15$\%, respectively. In contrast, the distributions of
simulated data of 2000 (Fig.~\ref{figure:SIMrednoise2} right 
panels) are totally inconsistent with the observed values
(P$<0.1$\% for both $\langle\alpha_{00}\rangle$ and $\sigma_{\alpha,00}$
distributions).

\begin{figure*}[hbt]
\noindent
\psfig{figure=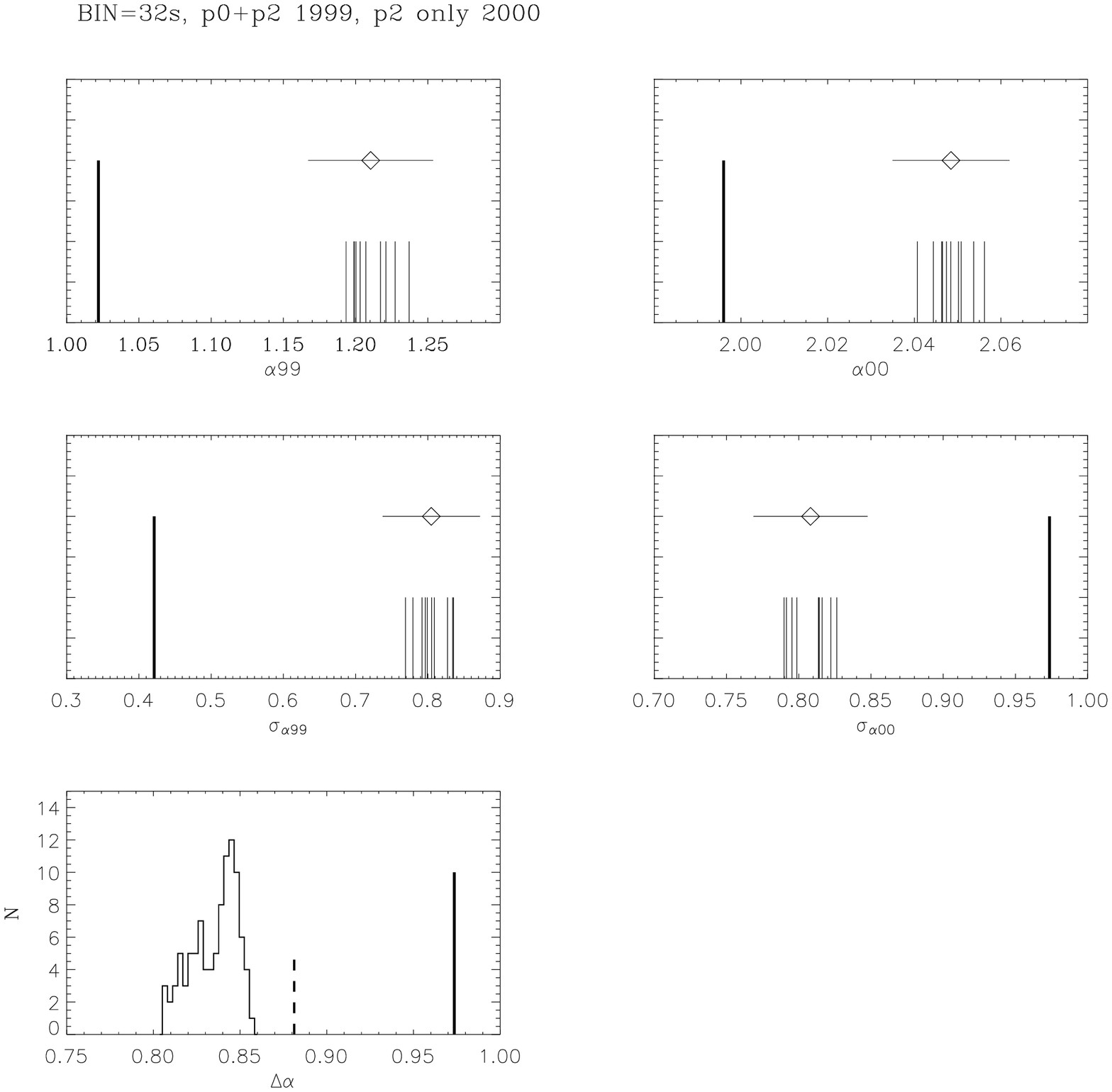,height=3.5cm,
bbllx=6pt,bblly=60pt,bburx=340pt,bbury=235pt,angle=0,clip=}
\psfig{figure=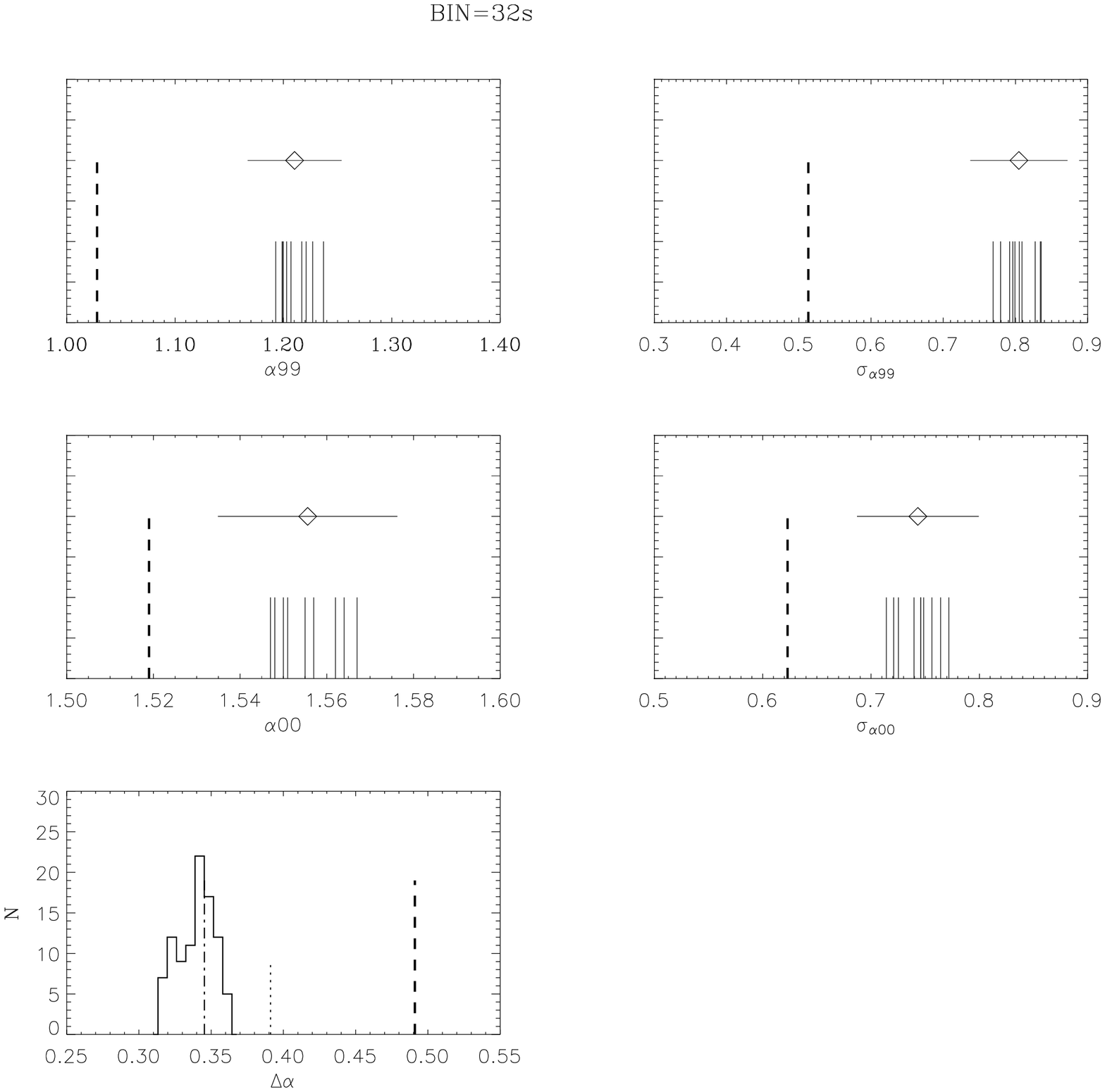,height=3.5cm,
bbllx=30pt,bblly=438pt,bburx=605pt,bbury=615pt,angle=0,clip=}
\caption{Left panel: Distribution of $\Delta\alpha$ obtained by shuffling the
time order of the \rxte\ orbits. The thick solid line is the location of
$\Delta\alpha_{\rm obs}$.
Middle panel: Distribution of $\langle\alpha\rangle$ (short, solid lines) 
obtained by shuffling the orbits order in 1999 (tbin=32 s). The open diamond gives the mean value
of the shuffled data and its error bar represents the 3$\sigma$ confidence level.
The long, dashed line represents the original value, obtained without any shuffling. 
The same symbol convention is used for the right panel, showing the
distribution of $\sigma_\alpha$ in 1999.
}
\label{figure:shuffle2}
\end{figure*}

The fact that $\Delta\alpha_{\rm synth}$ and $\Delta\sigma_{\rm synth}$ are not
centered around zero, because of the presence of different Poisson noise
components in the two light curves, motivated us to investigate further the the
impact of the Poisson noise on the distribution of the scaling index values. 
In principle, the effects of Poisson noise have been already accounted by the
red-noise simulations, where the ``appropriate" noise level has been computed
also considering the different background levels during 1999 and 2000.
Nevertheless, we considered what the effects on the SIM results would be if we
add, arbitrarily, some artificial extra Poisson noise (equal to half of the
noise actually present in the data)  to the synthetic light curves.  In
other words, we consider the possibility that the ``real" instrumental noise in
the light curves is, for some unknown reason,  underestimated in the previous
simulations, and what effects this may have on the SIM results.

The results of the noisy simulations are plotted in
Fig.~\ref{figure:SIMrednoise2b}.
Based on the $\Delta\alpha_{synth}$ distribution only  (top left panel), we
would conclude that the simulated results are in agreement
with  the observed value of $\Delta\alpha$ (the chance probability is indeed
P$\sim17$\%).  This might lead to the erroneous conclusion that the differences
observed using SIM  could be simply ascribed to the fact that the Poisson noise in the
previous simulations was not accounted for correctly. 
However, a look at the $\Delta\sigma_{synth}$ distribution  (Fig.~\ref{figure:SIMrednoise2b}, bottom left panel; P$<0.1$\%) reveals that this is not
the case. The same conclusion is reached looking at the synthetic distributions in
1999 and 2000, and keeping in mind that, in order to infer consistency between 
simulated and observed values, the distributions 
of {\it both} scaling index centroids {\it and} standard deviations should be consistent with the
respective observed values. 
This is clearly not the case in 1999 and in 2000: in the first case,
the synthetic distribution of $\langle\alpha_{99}\rangle$ is very poorly consistent 
with the observed value (P$=0.1$\%), whereas in 2000, it is the synthetic distribution of
$\sigma_{\alpha_{00}}$ values which is entirely 
inconsistent with $\sigma_{\alpha_{00,\rm obs}}$ (P$<0.1$\%).
 
Another macroscopic consequence of the arbitrary increase of the Poisson noise
is the the substantial right-ward shift of $\langle\alpha\rangle$. This  might
appear at odds with the lack of shift in $\langle\alpha\rangle$ observed in the
real data when passing from tbin=160 s to tbin=32 s (see
Fig.~\ref{figure:alpha}). However, this apparent discrepancy can be easily
explained by the most important property of the SIM, i.e. its ability to
discern an underlying structure in noisy data, {\it if} an underlying structure
is present. On the other hand, if no underlying structure is present (as in the
case of the simulated red-noise data), the increase of noise (and hence of
randomness) simply causes a shift of $\langle\alpha\rangle$ to larger values.

An alternative and {\it model-independent} way to assess if
$\Delta\alpha_{\rm obs}$ is dominated by Poisson noise effects is based
on the shuffling of the temporal order of the \rxte\ orbits. In this way,
the original long-term structure is destroyed, while the Poisson noise
levels and all the statistical moments of the distributions in 1999 and
2000 are unaffected.  If the orbit-shuffled light curves yield a
$\Delta\alpha$ consistent with $\Delta\alpha_{\rm obs}$, then we should
conclude that the difference in Poisson noise between the 1999 and 2000
(along with the difference in the short-term intra-orbit variability) is
the simplest explanation of the difference in scaling index observed. On
the other hand, if the values of $\Delta\alpha_{\rm shuffled}$ are not
consistent with the observed value, the direct consequence is that the
long-term temporal structure of the two original light curves does play
an important role.

The results of this test, carried out using light curves with tbin=32 s
because of their higher Poisson noise level (but similar results are
obtained with tbin=160 s), are shown in Fig.~\ref{figure:shuffle2}. The
left panel clearly shows that the orbit-shuffled results are inconsistent
with the $\Delta\alpha_{\rm obs}$ at very high significance level.  The
middle and right panels show the distributions of the centroid
$\langle\alpha\rangle$ and the width $\sigma_\alpha$ of the scaling index
histograms obtained by shuffling the orbits order in 1999 (very similar
results are obtained with 2000 data). Both plots show that the
orbit-shuffled quantities narrowly cluster at a location, which is,
statistically speaking, far away from the real data, indicating that the
long-term temporal structure contributes substantially to the SIM
results.

We conclude that, contrary to the results based on the comparison
between the 1999 and 2000 SFs and PDFs, the observed differences between
the respective distributions of the scaling index values are almost
certainly intrinsic, i.e. they cannot be explained by Poisson noise,
red-noise, nor any other instrumental effect that we can think of. To the
extend that the differences of the
scaling index histograms correspond to differences in the temporal
properties of the two light curves, we consider them as significant
evidence of non-stationarity in the the X-ray emitting process in
3C~390.3.

\section{Spectral Variability}
After searching systematically for the evidence of
temporal changes in 3C~390.3, we now want to investigate whether the 
apparent changes of the statistical 
timing properties are accompanied by spectral variations, and what is
the nature of the spectral changes (e.g., is 3C~390.3
experiencing an actual state transition?). To this end, we have performed two
different kinds of analysis.

\noindent

\begin{figure}
\psfig{figure=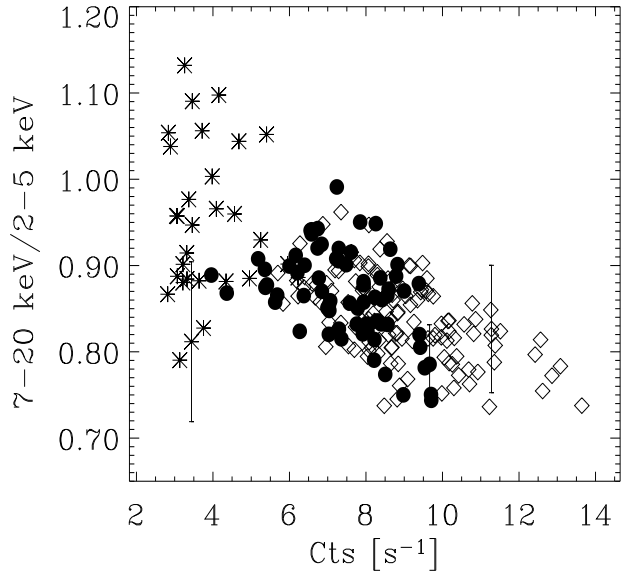,height=8cm,width=8.7cm,%
bbllx=70pt,bblly=60pt,bburx=410pt,bbury=350pt,angle=0,clip=}
\caption{7--20 keV/2--5 keV X-ray color plotted versus the
total count rate of \object{3C~390.3}. Star symbols correspond to the 
first 200 days of the 1999 light curve,
the filled dots describe the second part of the 1999 campaign, and the
open diamond the 2000 light curve. For the sake of clarity, only
a characteristic error for each cluster of points has been plotted. All the
errors associated with the X-ray colors  are
reported in Fig.~\ref{figure:lc} bottom panel. 
Time bins are 5760s ($\sim$
1 RXTE orbit).}
\label{figure:HR}
\end{figure}

1) The first (model-independent) method is based on the X-ray color --
count rate plot, often used for GBHs to separate the spectral states. The
plot of X-ray color (i.e., the ratio of the 7--20 over the 2--5 keV count
rate) versus the total 2--20 keV count rate is shown in
Fig.~\ref{figure:HR}. The hardness ratio clearly decreases with increasing
flux, during the second part of the 1999 and the whole of the 2000
campaigns. This trend is typical of type 1 radio-quiet AGN, and has been
observed many times in the past (e.g., Papadakis et al. 2002; Taylor et
al. 2004, and references therein).  This supports the conclusions of
Gliozzi et al. (2002), that the X-ray emission in 3C~390.3 is not
jet-dominated.

During the first half of the 1999 campaign, the source was at its lowest
flux state. The spectrum of the source also appears to be quite hard,
consistent with the general trend of spectral hardening towards low flux
states. However, the scatter of the hardness ratio points is very large,
and it seems that during this state, the spectral variations are more or
less independent of the source flux variations. So, based on the
color--count rate plot, it seems possible that during this period the
X-ray emission mechanism may behave in a different way, although the
errors associated with the X-ray colors (see Fig.~\ref{figure:lc} bottom
panel) prevent us from reaching a firm conclusion.

\begin{table}[ht] 
\caption{Time Intervals Used for Time-Resolved Spectral Analysis}
\begin{center}
\begin{tabular}{ccccc}
\hline
\hline
\noalign{\smallskip}
Start Time  & End Time  & Exposure & Flux${\rm ^a}$ & $\Gamma$ \\
(y/m/d h:m)&(y/m/d h:m)&(ks)& & \\
\noalign{\smallskip}       
\hline
\noalign{\smallskip}
\noalign{\smallskip}
99/01/08 00:26 & 99/02/25 17:58 & 17.90 & 1.87 &$1.55_{0.05}^{0.08}$\\
99/02/28 16:48 & 99/05/11 01:05 & 24.10 & 1.68 &$1.54_{0.05}^{0.03}$\\
99/05/14 00:58 & 99/07/10 20:15 & 21.20 & 2.64 & $1.64_{0.04}^{0.03}$\\
99/07/13 08:56 & 99/09/11 16:03 & 22.56 & 3.86 &$1.68_{0.03}^{0.02}$\\
99/09/14 18:07 & 99/11/13 22:58 & 23.38 & 3.71 &$1.67_{0.02}^{0.02}$ \\
99/11/17 14:25 & 00/01/03 08:00 & 18.69 & 4.71 &$1.74_{0.02}^{0.02}$\\
00/01/06 11:50 & 00/02/29 06:57 & 15.26 & 3.84  &$1.66_{0.03}^{0.02}$\\
00/03/03 04:26 & 00/05/11 12:43 & 32.46 & 4.90 &$1.70_{0.02}^{0.01}$\\
00/05/14 08:41 & 00/07/04 01:22 & 22.50 & 3.80  &$1.61_{0.02}^{0.03}$\\
00/07/07 00:09 & 00/09/01 22:29 & 26.91  & 4.10  &$1.63_{0.02}^{0.02}$\\
00/09/05 5:52 & 00/11/10 12:12 & 32.22  & 3.91   &$1.62_{0.02}^{0.01}$\\
00/11/13 12:57 & 00/12/31 03:12 & 25.47  & 5.03  &$1.67_{0.02}^{0.02}$\\
01/01/03 08:53 & 01/02/24 00:26 & 24.45  & 3.46  &$1.61_{0.02}^{0.02}$\\
\hline
\end{tabular}
\end{center}
${\rm ^a}$ 
2--10 keV flux in units of $(10^{-11}{\rm erg~ cm^{-2} ~s^{-1}})$, calculated assuming a
power law plus Gaussian line model.
\label{table:spec}
\end{table}

\begin{figure}
\psfig{figure=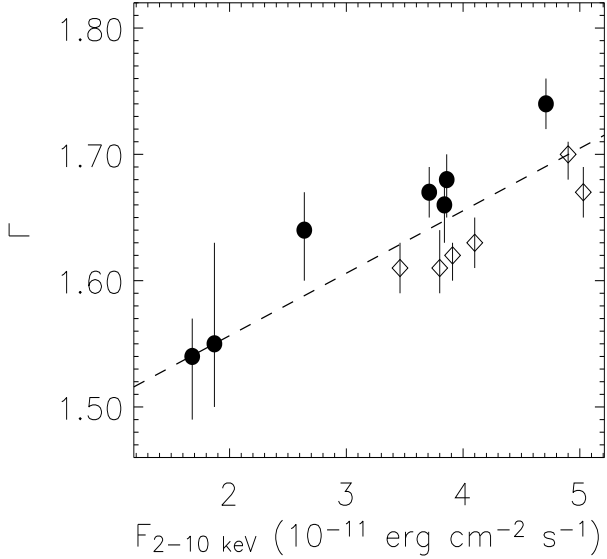,height=8cm,width=8.7cm,%
bbllx=70pt,bblly=60pt,bburx=385pt,bbury=325pt,angle=0,clip=}
\caption{Photon index, $\Gamma$, plotted versus the 2--10 keV flux
($10^{-11}{~\rm erg~cm^{-2}~s^{-1}}$).
Filled dots refers to the 1999 campaign, and the
open diamond to the 2000 light curve. The dashed line represents
a linear least squares fit to all the data points.}
\label{figure:gaflux}
\end{figure}

\noindent 2) More stringent constraints can be derived by using a second
(model-dependent) method: the time-resolved spectral analysis. Since the
data consist of short snapshots spanning a long temporal baseline, they
are well suited for monitoring the spectral variability of the source.

We divided the light curve of Figure~\ref{figure:lc} into thirteen
intervals, of duration 50--60 days each, with net exposures times ranging
between 15 ks and 32 ks. This scheme represents a trade-off between the
necessity to isolate intervals with different but well defined average
count rates and the need to reach a signal-to-noise ratio (S/N) high
enough for a meaningful spectral analysis.  The dates, exposure times and
mean fluxes in the thirteen selected intervals are listed in
Table~~\ref{table:spec}.

We fitted the 4--20 keV spectra with a model consisting of a power-law,
absorbed by the Galactic absorbing column, plus a Gaussian \feka\ line
with a rest energy fixed at 6.4 keV. In all the intervals, the line is
statistically required at more than 99\% confidence level, according to
an F-test. However, due to the low S/N, the line parameters are
characterized by large errors ($\sim$ 100\% on the line flux and width)
and thus their evolution in time cannot be investigated properly. On the
other hand, the photon indices are well constrained with uncertainties of
the order of 1--3\%. Figure~\ref{figure:gaflux} shows the plot of the
photon index $\Gamma$ versus the 2--10 keV flux, with the filled dots
representing intervals during the 1999 campaign and open diamonds during
the 2000 campaign. The photon index increases steadily with the 2--10 keV
flux, without any indication of a discontinuity between the $\Gamma$
corresponding to the low state and the other photon indices.

Based on this analysis, we conclude that 3C~390.3 is not undergoing
a transition between two distinct spectral states during the \rxte\ '99
and '00 monitoring campaigns. 

\section{Discussion}

We have studied two one-year long, well sampled X-ray light curves of
3C~390.3 from the \rxte\ archive, in search of evidence for
variations in its temporal properties. Our motivation stems from the fact
that a long-term \rosat\ light curve of the this broad-line radio galaxy
had shown evidence for non-linearity and possibly non-stationarity in the
past (Leighly \& O'Brien, 1997). The light curves we have used are the
longest, and best sampled light curves on time-scales of days/months of
this source to date. Hence, they offer a good opportunity to investigate
the important issue of stationarity in the X-ray light curves of AGN.

\subsection{Multi-technique Timing Approach: Linear Methods}

In order to assess whether the light curve of 3C~390.3 does show
variations in its temporal properties, we performed a thorough temporal
analysis based on different complementary timing techniques and extensive
Monte Carlo simulations. The first advantage of this approach is that it
leads to more robust results, not biased by the specific characteristics
of a single method. In addition, since the diverse methods employed probe
different statistical properties, they provide us with complementary
pieces of information that can be combined to constrain the physical
mechanism underlying the observed variability.

The {\it Power Spectral Density analysis} results suggest a possible shift of a
frequency break between 1999 and 2000, with $t_{\rm break}$ moving from $\sim$
1/300 -- 1/400 days to $\sim$ 1/50 --1/100 days.   However, this apparent shift
is not of high statistical significance.  The {\it Structure Function analysis}
also indicates a similar change in the characteristic time-scale of the
system.  Both SFs have a power-law shape up to lags shorter than $\sim 60$
days.  At longer time-scales, the 2000 SF appears to flatten to a roughly
constant level, while the 1999 SF keeps rising up to the largest time lags
sampled. Finally, the comparison between the two {\it Probability Density
Functions} (see Fig.~\ref{figure:PDF}), suggest the existence of an intrinsic
difference between the two time series, with the 1999 data characterized by a
bimodal distribution and the 2000 data following a uniform distribution.

However, based on the results from Monte Carlo red-noise simulations, which
were employed in order to test the significance of the apparent differences
between the 1999 and 2000 SF and PDFs, we found that they can be the result of
the red-noise character of the variability process with a probability at least
as high as $\sim 10$\%.

\subsection{The Scaling Index Method}

In contrast with the linear methods, that showed only suggestive evidence of a
change in the temporal properties of 3C~390.3 but were unable to rule out a
major influence of red-noise effects, the {\it Scaling Index ~method} clearly
reveals the presence of intrinsic changes that cannot by any means be explained
by red-noise nor Poisson effects (P$<$0.1\%). 

It is difficult to understand the consequences of this result at this
point. One obvious possibility is  that the variability process of \3c\ holds
more information than the red-noise PSD reveals. The main property  of the
synthetic light curves that we have produced using the method of Timmer \&
Koening (1995) is that their PSD has a specific red-noise shape.  The
disagreement then between the results from the simulated light curves and the
observed SIM values could imply that there is ``something" more than just ``red
plus Poisson noise" in the observed  light curves. In fact, there have been
indications the last few years that this may well be the case. For example,
Uttley, McHardy \& Vaughan (2004) have shown that  the rms-flux relation
recently discovered in the X-ray light curves of  AGN and GBHs implies that
their light curves are intrinsically ``non-linear". The simulations that we
have used in this work to test the SIM results do not reproduce this rms-flux
relation.  Perhaps then, the X-ray light curve of \3c\ is non-linear as well,
and the differences that we detect with the SIM method reflect this property.
Advanced simulations (which are beyond the scope of the present work) are
needed in order to investigate this issue further, and examine whether
non-linear synthetic light curves which do reproduce the rms-flux relation are
consistent with our results. 

On the other hand, it is also possible that, irrespective of whether the light
curves are linear or non-linear, the SIM differences that we observe imply real
differences in the temporal properties  of the source between 1999 and 2000.
The important issue of course is to identify these properties, in order to
advance our understanding of the variability mechanism. This is not an easy
task, because the nature of the AGN time variability is still poorly known
(apart from the ``linear vs. non-linear" issue, one has to consider other
issues as well like ``stochastic vs. chaotic" for example).  At the moment, no
clear physical insights can be derived from this non-linear tool.  Indeed, as
pointed out by Vio et al. (1992), a naive application and interpretation of the
non-linear methods results may lead to conceptual errors. 

Nevertheless, the much narrower 1999 scaling index distribution, with its
centroid $\langle\alpha\rangle$ located at a value significantly lower than
$\langle\alpha_{00}\rangle$, is consistent with the assumption that the 1999
time series is characterized by a ``higher degree of correlation". Since the
correlation structure of a time series is characterized by its auto-covariance
function, perhaps this result does suggest that the the width of the
auto-covariance function has decreased during the 2000 observations. This is
qualitatively in agreement, with the hypothesis that the characteristic
time-scale during 1999 is longer than during 2000, as suggested by PSD and SF
analyses. However, at this point this is merely a suggestion and alternative
hypotheses cannot be ruled out. If true, the differences in the scaling
indices do not challenge the ``red plus Poisson noise" view; they simply
demonstrate that the first term in this expression may vary with time.

\subsection{3C~390.3 and the analogy between AGN and GBHs}

It has been suggested many times in the past, and is now widely
believed, that AGN are simply the high mass analogs
of GBHs. For example, PSD analyses of long, high signal-to-noise, well
sampled X-ray light curves of a few AGN have shown clearly that their
power spectra are very similar to those of GBHs (e.g.  Uttley et al.,
2002; Markowitz et al., 2003; McHardy et al., 2004).

One of the most interesting aspects of the X-ray variability behavior of
the GBHs is that they often show transitions to different ``states''.
``State'' transitions have not been observed so far in AGN. The energy
spectral results reported in Section 5 clearly argue against the presence
of a bona fide state transition in 3C~390.3 as well. This is not
unexpected if we consider the time-scales at play: in GBHs state
transitions take place on time-scales of days (e.g. Zdziarski et al. 2004)
which translate into thousands of years for 3C~390.3, assuming a linear
scaling between the black hole masses.

However, even within the same state, the timing properties of GBHs, like
the PSD characteristic time-scales, vary with time. For example, during
the ``low/hard state'', the characteristic time-scales in the X-ray power
spectrum of Cyg X-1 change with time in a correlated way with the
spectral variations of the source (Belloni \& Hasinger, 1990; Pottschmidt
et al., 2003).  This is one example of ``stationarity-loss'', which can
offer important clues on the mechanism responsible for the observed
variations. These variations of the temporal properties, can take place
on time-scales as short as tens of seconds (e.g.  Casella et al. 2004).
These time-scales are consistent with the length of the 1999 and 2000
observations of 3C~390.3 which we study in this work, assuming again,
that characteristic time-scales scale linearly with black hole mass.

We find that the energy spectrum of the source in the first part of
the 1999 light curve is significantly harder than that of the high flux
peaks after day $\sim 400$ (Fig.~1).  Furthermore, the SIM results shows
variations in the timing properties of the source during the same
periods. In itself, this result strengthens the analogy between the X-ray
variability properties of AGN and GBHs: just like in the Galactic
systems, 3C~390.3 does show simultaneous spectral and intrinsic timing
property variations.

However, as we discussed above, the SIM results are not so easy to
interpret at the moment. On the other hand, the results presented in
sections 4.1-4.3, do suggest a much closer analogy between 3C~390.3 and
say Cyg X--1, but they are {\it not} statistically significant, possibly
because the quality of the light curves at hand is not good enough.

Our results show that the spectral slope, $\Gamma$, was $\sim 1.55$ during the
first $\sim 150$ days of the 1999 monitoring observations, increasing by a
factor of $\sim 10$\%, during the peak flux periods of the 2000 monitoring
observations. The Cyg X-1 slope does not become as hard as 1.55, nevertheless,
according to the results of Pottschmidt et al. (2003), when $\Gamma$ increases
from its lowest value of 1.8 to a value of $\sim 1.95-2$ (an increase by a
factor of $\sim 10$\% as well)   the centroid frequency of the $L_{2}$
component in its power spectrum increases by a factor of $\sim 5-6$ (as shown
in the middle panel of their Fig.~7). Since the $L_{2}$ centroid frequency is
commonly thought to correspond to the ``high-frequency"  PSD breaks in AGN,
this is the same  behavior that the results of 3C~390.3 {\it suggest}.

As mentioned above, we believe that the temporal and spectral behavior of
3C~390.3, as seen by \rxte\ during the 1999 and 2000 monitoring
campaigns, qualitatively mimics the properties of the GBH Cyg X--1 during
the low/hard state, provided that the appropriate time-scale differences
are taken into account.

\section{Summary and Conclusions}
The main results and conclusions of this work can be summarized as follows:
\begin{itemize}

\item We have studied in detail the variability characteristics of the
two-year long, \rxte\ monitoring light curve of \3c. Linear methods, like
PSD, SF and PDF analysis show suggestive evidence of a change in the
characteristic time-scale of the system between 1999 and 2000. 

\item However, only the results from the scaling
index method show a clear and significant evidence of a change in the intrinsic
properties of the system that cannot be explained by red or Poisson noise
effects. 

\item It is not easy to interpret the SIM results at this moment. To the extend
that the scaling index is related with the timing properties of the variability
process,  one possibility is a  change of the characteristic time-scale between
1999 and 2000, in agreement with the suggestive results from the PSD, SF and
PDF analyses.

\item Our results reinforce the similarity between the X-ray variability
properties of AGN and GBHs. In analogy with GBH phenomenology, 
the loss of stationarity in 3C~390.3 seems to correspond
to an increase in the frequency break as the spectrum softens, as
observed in Cyg X-1, when the system is in its low/hard spectral state.

\item  Since the SIM is not yet a standard technique, further work is needed in
order to investigate its properties and its possible relation with more
frequently used (and better understood) statistical analysis tools like the power spectrum. 
Nevertheless, we believe
that the present results demonstrate the benefits from the application of the
SIM (and possibly other non-linear methods) in the analysis of the AGN X-ray 
light curves. 

\item  The \rxte\ archive is
populated with numerous, long AGN light curves. This is a good opportunity to
start a systematic study of these data, for a search of spectrally related
variations of the timing properties of these systems. Since the outcome of any
Monte Carlo simulation based on an assumed ``model-PSD'' is inevitably dependent
on the assumptions made, we believe that a thorough analysis of a large number 
of high-quality AGN light curves is the most direct way
to examine whether the present results are indicative of the AGN
behavior in general, or just a peculiarity associated with \3c. 
We plan to explore this issue in a future work.

\end{itemize}

\begin{acknowledgements}
We would like to thank P. Uttley, S. Vaughan and A. Markowitz for the lengthy
discussions and their criticisms of the original draft of this paper, and the 
methodology used herein. We would also like to thank the anonymous referee for the useful
comments and suggestions that improved the paper. We are grateful to H.
Scheingraber for providing the SIM code. Financial support from NASA LTSA
grants NAG5-10708 (MG) is gratefully acknowledged. 

\end{acknowledgements}

\end{document}